\documentclass[twocolumn,showpacs,preprintnumbers,amsmath,amssymb,prb]{revtex4}

\usepackage{graphicx}
\usepackage{dcolumn}
\usepackage{bm}
\usepackage{amsmath}
\usepackage{placeins}
\usepackage{textcomp}
\usepackage{bbold}
\usepackage{xcolor}
\newcommand{\ra}{\rightarrow}

\newcommand{\bracket}[1]{\left\langle #1\right\rangle}

\begin{document}


\title{Counting statistics of hole transfer in a p-type GaAs quantum dot {with dense excitation spectrum}}

\author{Y. Komijani\footnote{Now at Department of Physics and Astronomy, University of British Columbia, Vancouver, B.C., Canada V6T 1Z1}}
\email{komijani@phys.ethz.ch}
\author{T. Choi}
\author{F. Nichele}
\author{K. Ensslin}
\author{T. Ihn}
\affiliation{Solid State Physics Laboratory, ETH Zurich, 8093 Zurich, Switzerland}
\author{D. Reuter\footnote{Now at University Paderborn, Department Physik, Warburger Stra\ss e 100, 33098 Paderborn, Germany}}
\author{A. D. Wieck}
\affiliation{Angewandte Festk\"orperphysik, Ruhr-Universit\"at Bochum, 44780 Bochum, Germany}

\date{\today}

\begin{abstract}

Low-temperature transport experiments on a p-type GaAs quantum dot capacitively coupled to a quantum point contact are presented. The time-averaged as well as time-resolved detection of charging events of the dot are demonstrated and they are used to extract the tunnelling rates into and out of the quantum dot. The extracted rates exhibit a super-linear enhancement with the bias applied across the dot which is interpreted in terms of a dense spectrum of excited states contributing to the transport, characteristic for heavy hole systems. The full counting statistics of charge transfer events and the effect of back action is studied. The normal cumulants as well as the recently proposed factorial cumulants are calculated and discussed in view of their importance for interacting systems.

\end{abstract}

\pacs{73.23.-b, 73.63.-b}
\maketitle


\section{Introduction}

Quantum dots (QDs or simply dots) are small conducting islands that confine charge carriers in three dimensions resulting in a discrete spectrum of excited states. This spectrum is often studied in transport experiments by measuring the current,\cite{Kouwenhoven97} that is allowing carriers to tunnel between the dot and source and drain leads (reservoirs). The capacitive coupling of QD to nearby gates enables tuning the energy of excited states with respect to electrochemical potential of the leads. It is a fascinating experimental observation that a similar capacitive coupling to a nearby electrical current passing through a constriction provides the possibility of measuring the charge of the dot with a precision of a small fraction of an electron's charge.~\cite{Field93} The conductance of the constriction changes as a function of the average charge population of the QD.

Two-level fluctuations (random telegraph noise) of the detector current provide more information about the dot than just the average current. The fluctuations of the current enable the time-resolved detection of single-particle charging and de-charging events in the QD.~\cite{Vandersypen04,Elzerman04a,Schleser85} This, on the other hand, reveals more information about the energy spectrum of the dot, the relaxation of excited states to the ground state and their coupling to the leads.~\cite{Gustavsson09} This information is valuable for the case of p-type QDs where the present understanding of their properties is limited by the lack of experimental results.

Counting statistics of the charge transfer is another tool to study quantum dots. Experimental studies of counting statistics using charge detection with a quantum point contact (QPC) were started by Gustavsson et al.,~\cite{Gustavsson06b,Gustavsson07} and Fujisawa et al.,~\cite{Fujisawa06} and continued by Fricke et al.~\cite{Flindt09,Fricke10b} All these experiments were performed on n-type GaAs or InAs electronic systems.~\cite{Choi12} Therefore it is interesting to compare these results with those obtained in a QD realized on a  p-GaAs two dimensional hole gas (2DHG), where the carrier-carrier interactions are supposedly stronger both in the dot and in the leads compared to n-type systems.

In this article we investigate these effects in a p-type GaAs QD system for which heavy holes (HHs) are the main carriers. The large effective mass of holes ($m^*_{\rm HH}\sim 0.4m_0$~\cite{Winkler03,Fabrizio}) is several times larger than that of conduction band electrons making carrier-carrier interaction effects more pronounced compared to the kinetic energy than in their electronic counterparts.~\cite{Komijani10} The same reason leads to the fact that screening is expected to be stronger and that the single-particle energy spacing is much smaller in confined p-type systems, making it very difficult to be resolved at accessible temperatures.~\cite{Komijani08} Additionally, the strong spin-orbit interaction in the valence band holds promise for interesting spin physics in these QDs.~\cite{Bulaev07} Successful optical manipulation of holes and large coherence times measured in these experiments (an order of magnitude larger than electrons)~\cite{Brunner09,Eble09,Fallahi10,Greilich11,Chekhovich13} is another motivation for realization of hole-based qubits and their studies using transport.

However, the fabrication of tunable p-type QDs is challenging,~\cite{Grbic05} essentially because metallic gates on top of shallow p-doped heterostructures have a low Schottky barrier resulting in leaky and hysteretic behavior {and different fabrication techniques have to be adopted}. The Coulomb blockade effect in lithographically-defined dots in p-type GaAs heterostructures was first demonstrated by Grbic et al.~\cite{Grbic05} using local oxidation lithography. The same technique was later shown to be effective in further confining the carriers and observing the individual excited energy states in a QD.~\cite{Komijani08} Induced SETs were also fabricated using undoped GaAs heterostructures and were shown to exhibit Coulomb blockade effects.~\cite{Klochan10} In spite of this progress the level of control on the fabrication of these nanostructures and the understanding of the role of interactions in hole systems is still far from complete. In this article we attempt to improve on this understanding by realizing time-resolved charge detection of hole tunnelling into a QD fabricated by shallow wet chemical etching.

\section{Sample and setup}

\begin{figure}
\includegraphics[width=0.45\textwidth,trim=0 0 0 2.1cm]{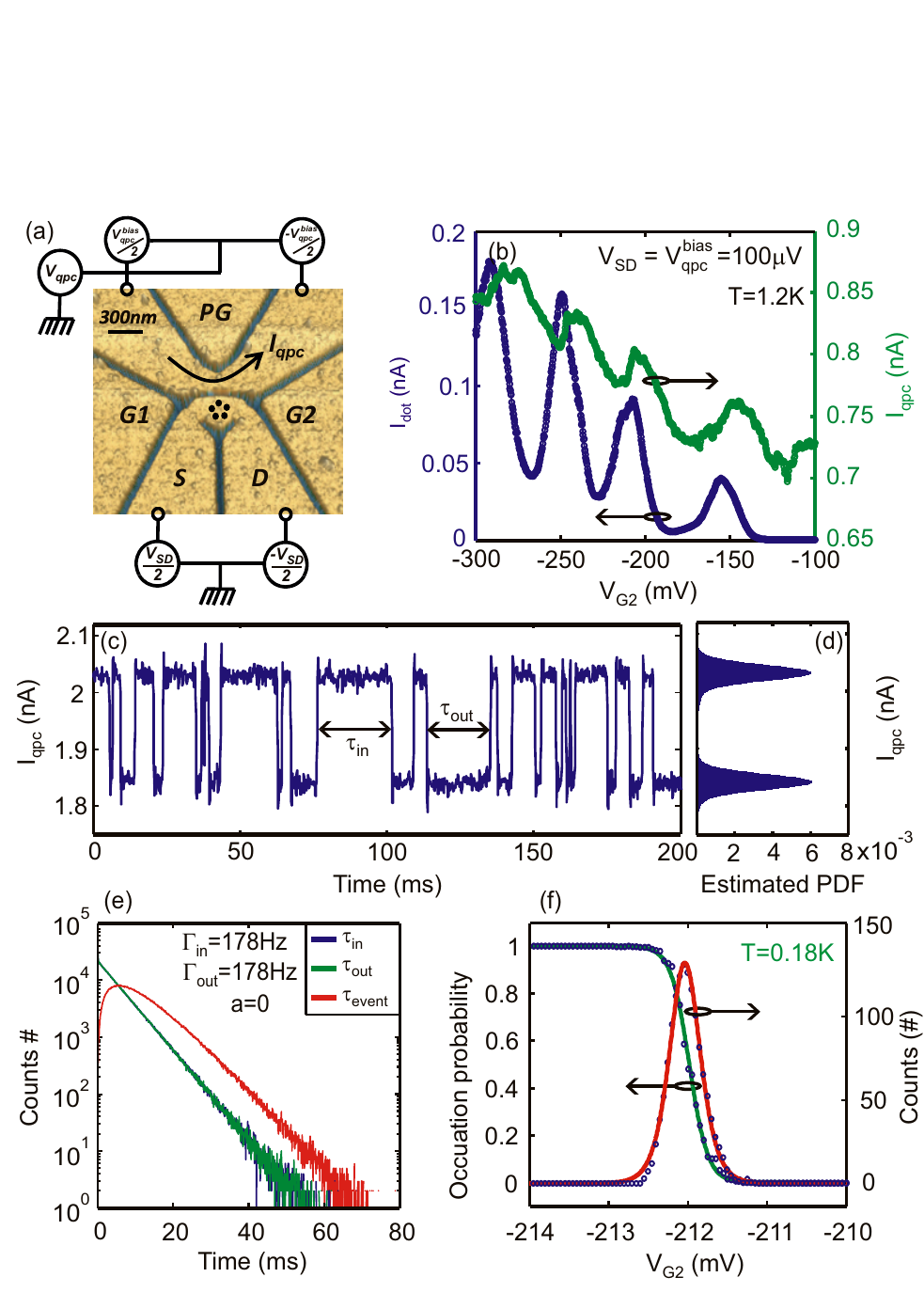} \caption{\label{fig:Fig1} \small{
(a) AFM micrograph of the sample consisting of a QD with an integrated charge read-out QPC. The dark regions are trenches created on the surface of the heterostructure by chemical etching. The applied voltages are shown on the same figure. (b) QD current (blue) and QPC current (green) as a function of $V_{\rm G2}$ measured at the temperature of $T\approx$~1.2\,K. (c) QPC current as a function of time, showing a few holes tunnelling into and out of the QD over a timescale of 200\,${\mu}$s. The lower current level corresponds to a state when the dot holds one excess hole. The QPC current was filtered with a 3\,kHz software filter and re-sampled at a frequency of 14\,kHz. The random variables $\tau_{\rm in}$ and $\tau_{\rm out}$ quantify the times it takes for a hole to tunnel into and out of the dot, respectively. (d) The probability density function (PDF) obtained from normalized histogram of the detector current showing two distinct current levels corresponding to the two charge states of the QD (e) The histogram of times a hole needs to tunnel into the QD (blue), tunnel out of the QD (green), and the event time defined as the sum of two consecutive tunnelling in and out events (red) for a symmetric configuration ($\Gamma_{\rm in}=\Gamma_{\rm out}$). The green curve exactly overlaps with the blue curve. (f) The occupation of the QD and the number of charge events as the gate voltage is swept over a Coulomb blockade peak with fits to the Fermi-Dirac distribution (green curve) and its derivative (red curve), respectively.}}
\end{figure}

Fig.\,\ref{fig:Fig1}(a) shows an AFM micrograph of the sample which was patterned in the 2DHG by electron beam lithography followed by shallow wet chemical etching.~\cite{Komijani10} The trenches seen in Fig.\,\ref{fig:Fig1}(a) are 20\,nm deep and locally deplete the 2DHG situated 45\,nm below the surface, thereby separating the 2DHG plane into laterally disconnected regions. Each of them is connected to metallic leads via ohmic contacts. The host material consists of a C-doped GaAs/AlGaAs heterostructure grown along the (100) plane.~\cite{Wieck00} Prior to sample fabrication the quality of the 2DHG was characterized by standard magnetotransport measurements at 4.2\,K and a hole density of $n$~=~2.7$\times$10$^{11}$\,cm$^{-2}$, and a mobility of $\mu$~=~60'000\,cm$^2$/Vs were obtained. 

The sample consists of a QD together with a nearby QPC. The measurement setup and the applied bias voltages are also schematically shown in Fig.\,\ref{fig:Fig1}(a). The dot bias and the QPC bias are both applied symmetrically. The overall potential of the QPC ($V_{qpc}$) is used to control the electrochemical potential of the QD, while the plunger gate (PG) is used to tune the QPC transmission. The in-plane gates G1 and G2 are used to tune the tunnel coupling between the QD and source (S) and drain (D), but they also have a significant lever arm on the dot.

\section{Results and discussion}


\subsection{Time-averaged/time-resolved charge detection}
Fig.\,\ref{fig:Fig1}(b) shows simultaneous measurements of QPC and QD currents as a function of the voltage applied to the gate G2 at the temperature $T\approx$~1.2\,K. As the gate voltage is increased, the holes are unloaded from the dot one by one. The dot current shows clear conductance resonances at the charge degeneracy points where the charge state of the dot changes by one elementary charge. This can be clearly seen as a step of 30\,pA in the QPC current ($\approx$ 4\%) at the position of the Coulomb peaks. Note that the average QPC current decreases with G2 (due to the corresponding lever arm), since no electrostatic compensation was performed here in order to avoid activating additional fluctuators in the sample which degrade the detector signal. For the remainder of the paper we will consider the results obtained in a dilution refrigerator with a base temperature of 100\,mK.

When the bandwidth $\Gamma_D$ of the detector circuit is small compared to the tunnelling rates of the QD, it only responds to the average charge population of the dot. As $\Gamma_D$ is increased to about 3\,kHz and the tunnelling barriers are tuned sufficiently opaque, time-resolved charge detection becomes possible. The detector current then exhibits a two-level fluctuating behavior as a function of time because holes tunnel into and out of the dot [shown in Fig.\,\ref{fig:Fig1}(c)]. The two levels on the histogram of the detector current [Fig.\,\ref{fig:Fig1}(d)] are a result of the two charging states of the dot. The random variables $\tau_{\rm in/out}$ quantify the time it takes for a hole to tunnel into or out of the QD and are used to calculate the tunnelling rates according to $\Gamma_{\rm in/out}=\langle \tau_{\rm in/out}\rangle^{-1}$ (Here bracket denotes an ensemble averaging). The latter can be used to quantify the coupling symmetry of the QD to the leads by defining the normalized coupling asymmetry $a=(\Gamma_{\rm in}-\Gamma_{\rm out})/(\Gamma_{\rm in}+\Gamma_{\rm out})$. The histograms of $\tau_{\rm in/out}$ and the event length ($\tau_{\rm event}=\tau_{\rm in}+\tau_{\rm out}$) are plotted in Fig.\,\ref{fig:Fig1}(e) for a symmetric configuration of the QD ($a=0$). About 2 million events, accumulated over more than 5 hours, were used to produce these histograms indicating the stability of the sample. The exponential distribution of the tunnelling rates motivates the use of the rate equation technique to study the statistics of the charge transfer. The short-time suppression of $\tau_{\rm{event}}$ is a consequence of the correlated transport and sequential tunnelling through the QD.~\cite{Gustavsson09}

Figure \ref{fig:Fig1}(f) shows the occupation probability of the QD (extracted from the duty cycle of the detector current) together with the average number of events (in 1 second time traces) as the gate voltage is swept over a Coulomb blockade peak. They can be fitted with a Fermi-Dirac distribution, from which a hole temperature of $T_{\rm hole}\sim$~180\,mK is obtained. 


\subsection{Excited state spectrum}
\begin{figure}
\includegraphics[width=0.45\textwidth]{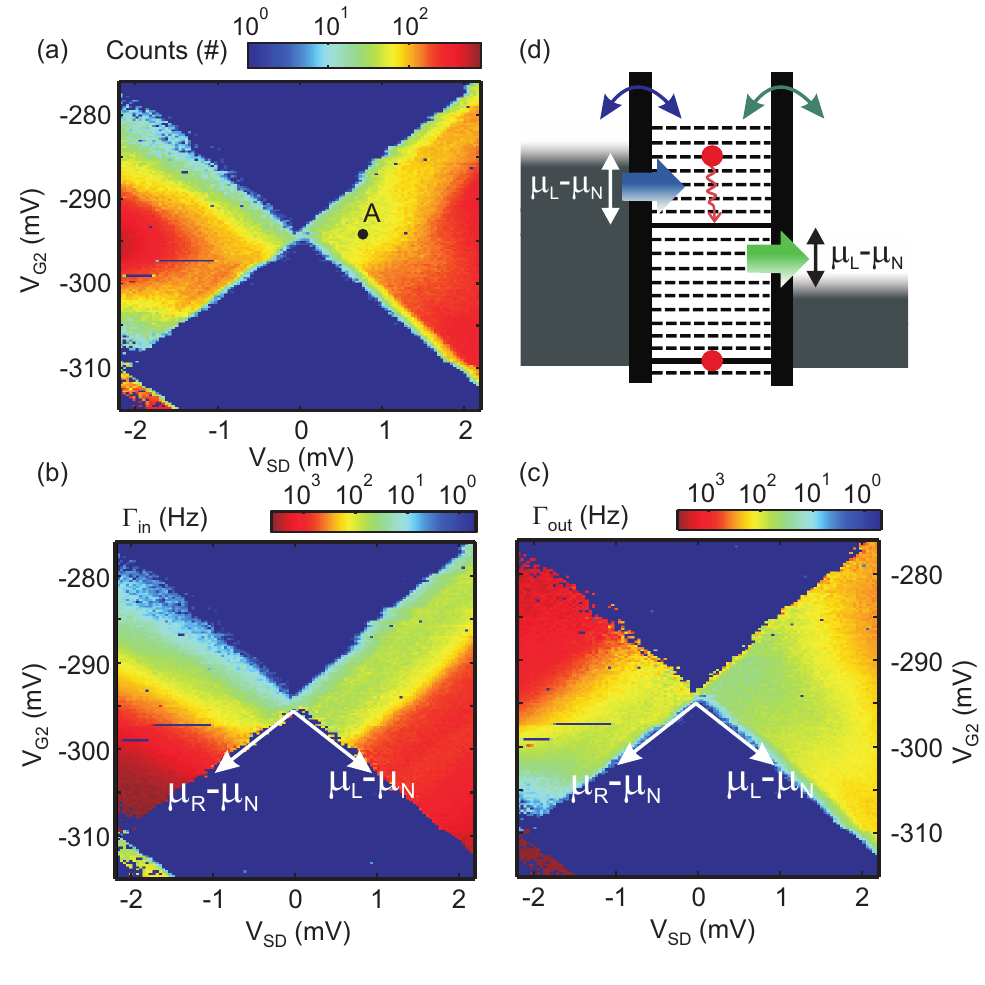} \caption{\label{fig:Fig2} \small{
(a) Event rate $\Gamma_{\rm event}$, and tunnelling rates (b) $\Gamma_{\rm in}$ (c) $\Gamma_{\rm out}$ as a function of the gate voltage and the applied QD bias demonstrating half of a Coulomb diamond. All the rates increase with the dot bias. White arrows indicate the relative position of the QD electrochemical potential $\mu_N^0$ with respect to those of source and drain. (d) In the case of a moderate energy-dependence of barriers, tunnelling into and out of the dot both involves many excited states and relaxation does not change the qualitative picture.}} \end{figure}

Fig.\,\ref{fig:Fig2}(a) shows the event rate as a function of the QD bias $V_{\rm SD}$ and the gate voltage $V_{\rm G2}$. Due to a strong energy-dependence of the tunnelling rates, tunnelling on the adjacent Coulomb peaks is either too fast or too slow to be properly detected. The charging energy of the QD is $E_C\approx$~2\,meV which corresponds to a total capacitance of $C_{\Sigma}\approx$~40\,aF. Assuming a disk-like shape for the dot with $C_{\Sigma}=4\varepsilon_0\varepsilon_r r$, where $r$ is the dot radius and $\varepsilon_r=12.9$ for GaAs, this provides an (upper) estimate of $\approx$~160\,nm for the electronic diameter of the dot, and an upper limit of 55 for the number of holes in the QD. With this diameter the mean single-particle level spacing can be calculated from $\Delta=\pi\hbar^2/m^*_{\rm HH}A$ with $A=\pi r^2$, giving $\Delta\approx$~29\,$\mu$eV comparable to $k_BT$. The large effective mass of the holes results in a dense spectrum of confined states and therefore p-type QDs can be considered to be in the crossover between electron QDs with a discrete, and metallic SETs with a continuous excited state spectrum for the sizes investigated here. This manifests itself in the fact that it is not possible to resolve excited states in the diamond measurements as shown in Fig.\,\ref{fig:Fig2}. Were this resolution possible, we would expect a stepwise increase of the number of events with the steps parallel to the edges of the diamond.~\cite{Gustavsson09} Nevertheless it can already be seen in this figure that the number of events generally increases with increasing QD bias. Note that the event rate in Fig.\,\ref{fig:Fig2}(a) is not symmetric with applied bias, presumably due to some tunnelling asymmetry of the barriers.

More insight into the role of the dot excitation spectrum is obtained by looking at Fig.\,\ref{fig:Fig2}{(b)-(c)} where $\Gamma_{\rm in}$ and $\Gamma_{\rm out}$ are plotted instead of the number of events. The increase in $\Gamma_{\rm in}$ and $\Gamma_{\rm out}$ with $V_{\rm SD}$ can be understood in terms of the additional available tunnelling channels in a system with dense spectrum. The relative position of the QD electrochemical potential $\mu_N^0$ with respect to those of left ($\mu_L$) and right ($\mu_R$) leads are indicated with white arrows. Lines of constant $\Gamma_{\rm in/out}$ are parallel to the edges of the diamonds. In particular, $\Gamma_{\rm in}$ depends only on the difference of the electrochemical potentials of the source ($\mu_S$) and the dot $\mu_{S}-\mu_N^0$ (for positive bias $\mu_S=\mu_L$ and $\mu_D=\mu_R$ while for negative bias $\mu_S=\mu_R$ and $\mu_D=\mu_L$). This suggests that the number of available (excited) states between these two levels is the cause of the increase in the tunnelling rate. Provided that it has enough energy, a tunnelling-in hole can occupy any of these states and this increases the tunnelling rate with bias (see the Appendix for a simple example in which tunnelling rate into the dot becomes the sum of tunnelling-in rates into ground state and excited state). Similarly, $\Gamma_{\rm out}$ depends only on the difference between the electrochemical potentials of the drain ($\mu_D$) and the dot $\mu_N^0-\mu_D$, meaning that the number of options for holes tunnelling-out also increases with the bias. For example, it is possible for a hole in the $(N+1)$-hole ground state to tunnel out and leave the dot in any of the $N$-hole excited states. Motivated by these ideas, the level diagram of the dot is represented in Fig.\,\ref{fig:Fig2}(d) with a dense ladder of excited states both \emph{above} and \emph{below} the ground state transition $\mu_N^0$ with an electrochemical potential of $\mu_{N}^{\pm m}$ (the index $m$ refers to a transition involving excited states) which contribute to $\Gamma_{\rm in}$ and $\Gamma_{\rm out}$, respectively.

\subsection{Rate equation simulation}
To verify the ideas discussed in the previous section we performed rate equation simulations, for a dot in which both $N$ and $(N+1)$-charge configurations have many excited states. The occupation probabilities ($p_i^{N}$ and $p_j^{N+1}$) of individual states are calculated in the steady state and the total tunnelling-in/out rates are obtained from
\begin{equation}
\begin{split}
\Gamma_{\rm out}&=\sum_{ij}{\Gamma_{\rm N \leftarrow N+1}^{i\leftarrow j}p_j^{N+1}}/\sum_j{p_j^{N+1}} \\
\Gamma_{\rm in}&=\sum_{ij}{\Gamma_{\rm N+1 \leftarrow N}^{j\leftarrow i}p_i^{N}}/\sum_i{p_i^{N}} \\
\end{split}
\end{equation}
The parameters of the model ($\bar f(\epsilon)\equiv 1-f(\epsilon)$)
\begin{eqnarray}
\Gamma_{\rm N\leftarrow N+1}^{i\leftarrow j}&=&\Gamma_L\bar f(\mu_L-\mu_{N}^{j-i})+\Gamma_R\bar f(\mu_R-\mu_{N}^{j-i}) \nonumber \\
\Gamma_{\rm N+1\leftarrow N}^{j\leftarrow i}&=&\Gamma_Lf(\mu_L-\mu_N^{j-i})+\Gamma_Rf(\mu_R-\mu_N^{j-i})\nonumber
\end{eqnarray}
depend on the gate and the applied bias through the argument of Fermi distributions $f(\epsilon)$. All the states in the dot are assumed to be coupled to the leads with the same coupling ($\Gamma_L=\Gamma_R=$~100\,Hz). Since we expect our QD to be far from the few-hole regime, a linear spectrum with a constant mean-level spacing is assumed: 20 levels with the energy-separation of 100\,$\mu$eV are taken into account. A strong energy relaxation ($\gamma=$~1\,kHz) to the ground state is assumed for all levels.

\begin{figure}
\includegraphics[width=0.45\textwidth]{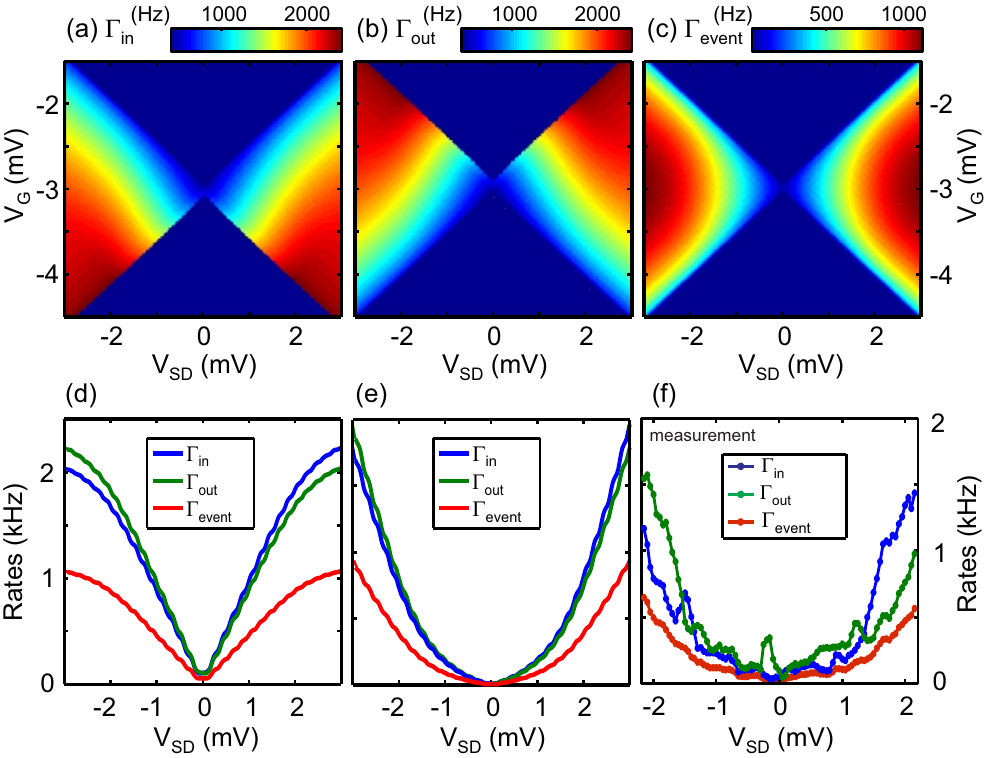} \caption{\label{fig:Fig3}\small{
Rate equation simulation of a dot with a dense spectrum of 100\,$\mu$eV equi-distant energy levels. $\Gamma_{\rm in}$ in (a) and $\Gamma_{\rm out}$ in (b) both increase with the dot bias. The equi-$\Gamma_{\rm in}$ lines and equi-$\Gamma_{\rm out}$ lines are parallel to source and drain lines, xrespectively. (c) $\Gamma_{\rm event}$. (d) Horizontal cut through $\Gamma_{\rm in}$, $\Gamma_{\rm out}$ and $\Gamma_{\rm event}$ in figures (a)-(c) at resonance. (e) Same as (d) but assuming an exponential energy-dependence of barriers. (f) Horizontal cut through the measurement data in Fig.\,\ref{fig:Fig2}(a)-(c) at resonance as in (d).}}
\end{figure}

The tunnelling-in rate $\Gamma_{\rm in}$ is shown in Fig.\,\ref{fig:Fig3}(a). It increases each time an excited-state transition of the $(N+1)$-charge configuration enters the bias window. Had we assumed no excited states for the $N$ charge configuration, $\Gamma_{\rm out}$ would stay constant and the event rate would saturates at the value of the tunnelling-out rate, which would become the bottleneck. 

However, since a similar spectrum of excited states is assumed for the $N$-charge configuration, $\Gamma_{\rm out}$ also increases with the bias and this simple model is able to qualitatively reproduce the measurement result, as shown in Fig.\,\ref{fig:Fig3}. Fig.\,\ref{fig:Fig3}(d) shows a horizontal cut through figures\,\ref{fig:Fig3}(a)-(c) at resonance. The linear increase of the tunnelling rates with bias is a consequence of equal tunnel couplings of individual states which is slightly different than the non-linear increase of the tunnelling rates observed in the measurement (Fig.\,\ref{fig:Fig3}(f)), but it must be noted that the difference can be easily captured by assuming an energy-dependence of the tunnelling rates as shown in Fig.\,\ref{fig:Fig3}(e).



\subsection{Counting statistics}
\begin{figure}
\includegraphics[width=0.45\textwidth]{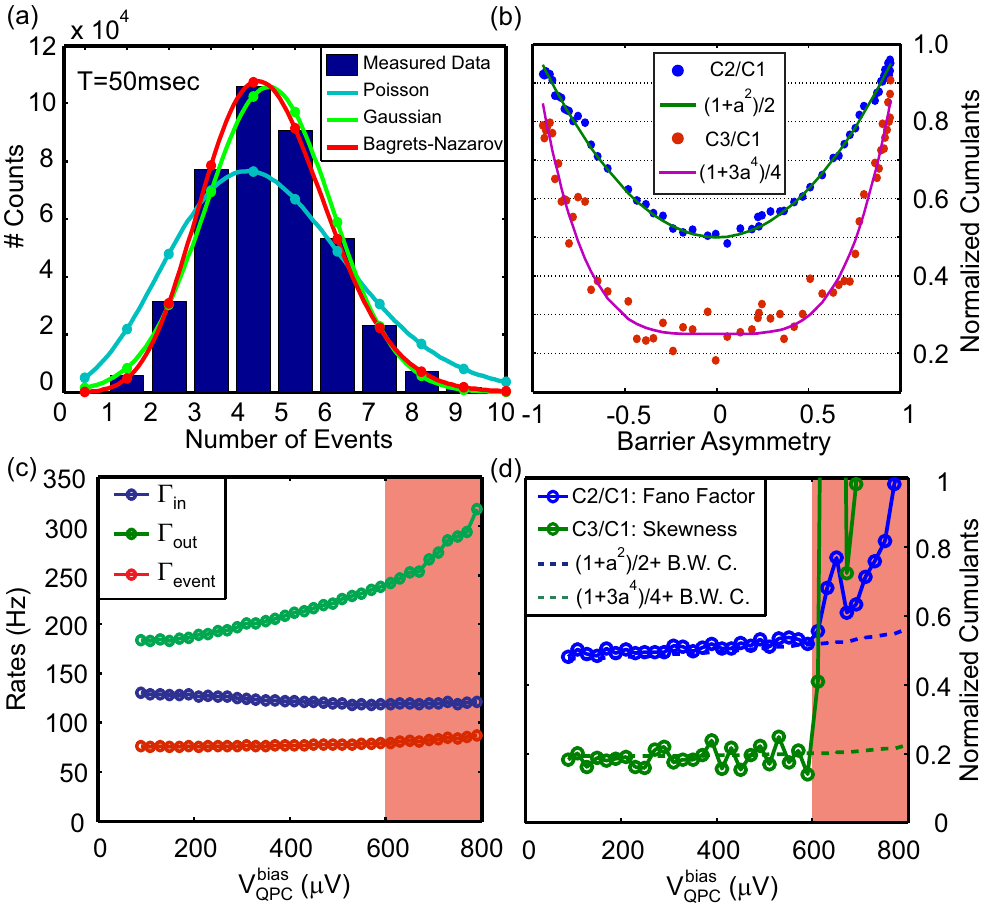} \caption{\label{fig:Fig4}\small{
(a) Histogram of the number of events during T=50\,msec at the point A in Fig.\,\ref{fig:Fig2}(a) with symmetric barriers $(a=0)$. The blue and green curves show the Poisson and Gaussian distributions respectively, calculated using the mean and variance of the measured data (all these distributions are discrete and the connecting lines are just guides to the eye). While the Gaussian distribution fits reasonably well to the histogram, the data is best described by the Bagrets-Nazarov distribution with the input parameters $\Gamma_{\rm in}$ and $\Gamma_{\rm out}$. (b)  The first two normalized cumulants of the statistical distribution of events as a function of tunnelling barrier asymmetry $a$, calculated from 100 time traces, each 10\,sec long. The blue points show $C2/C1$ or the Fano factor and the red points show $C3/C1$ which is the skewness. The solid lines are the model predictions. While the Fano factor agrees quite well with the model the skewness points are scattered much more due to limited statistics. (c) $\Gamma_{\rm in}$, $\Gamma_{\rm out}$ and $\Gamma_{\rm event}$ as a function of the bias voltage on the QPC. The decrease in $\Gamma_{\rm in}$ and increase in $\Gamma_{\rm out}$ is most probably due to a gating effect. (d) Fano factor and skewness as a function of QPC bias showing that the statistics of electron transport is not influenced by the emission of energy quanta by the QPC. The dashed lines show the model predictions discussed in the text augmented by finite-bandwidth corrections (B.W.C).~\cite{Naaman06} The red shaded area shows the onset of the detector signal degradation due to charge fluctuators in the QPC (Counting was not possible for $V_{\rm QPC}^{\rm bias}<50$\,$\mu$V).}}
\end{figure}

Fig.\,\ref{fig:Fig4}(a) shows the histogram of the number of events in a 50\,msec time window with symmetric tunnelling coupling ($a=0$) of the dot (point A in Fig.\,\ref{fig:Fig2}(a)). Three distributions, namely the Poisson distribution, the Gaussian  distribution and the model of Bagrets-Nazarov~\cite{Bagrets03} with two-states are plotted in the same figure. The fact that the model of Bagrets-Nazarov (BN) matches perfectly to the data indicates that, in spite of the dense spectrum of the dot and its contribution to transport (as shown in the previous section), the statistics is dominated by a two-state Markovian model. This is presumably due to a strong relaxation in the quantum dot. Using a simple model in the Appendix, it is shown that in presence of a single excited state, the visible 

The barrier asymmetry $a$ can be tuned in our experiment by applying asymmetric voltage offsets to the gates G1 and G2 while keeping their symmetric component constant in order to stay at the point A of Fig.\,\ref{fig:Fig2}. The Fano factor ($F=C_2/C_1$) and the skewness ($S=C_3/C_1$) extracted from the data are plotted as a function of the asymmetry $a$ in Fig.\,\ref{fig:Fig4}(b) together with the corresponding predictions of the BN model ($C_n$ is n-th cumulant of number of events in a series of 10\,sec-long time traces). The agreement of the model with the data indicates that again a two-state Markovian model is sufficient to describe the observed statistics.

It is not {\it a priori} clear whether each event in a given time trace corresponds to a charge transfer from source to the drain. In a quantum dot without any excited states in the relevant energy window, the ratio between the charges tunnelled back to the source to those transferred to the drain is essentially $k_BT/V_{SD}$ which is 1/40 at the point A in Fig.\,\ref{fig:Fig2}(a). This has motivated the use of counting experiments as an accurate tool to measure the current in this weakly-coupled regime, in which the current itself is too small to be directly measured.~\cite{Choi12}
However, in a quantum dot with a dense spectrum with energy-dependent tunnelling rates the ratio can be higher. In the presence of a strong energy-dependence of the tunnelling rates, relaxation (whose rate is denoted by $\gamma$) is crucial to ensure fully uni-directional transport. 
While this condition $\Gamma_{\rm in/out}\ll\gamma$ is presumably satisfied in our case ($\Gamma_{\rm in/out}\ll\Gamma_D\ll\gamma$) with the detector bandwidth of $\Gamma_D=$~3\,kHz, this might not be the case in more strongly coupled regimes.

%


\subsection{QPC back-action}

The power dissipated in and around the QPC is emitted as photons and phonons close to the QPC and hence may cause back-action on the QD either by increasing the effective temperature of the leads,~\cite{Gasser09} or by excitation of the QD due to photon and phonon-assisted tunnelling (PAT).~\cite{Kouwenhoven94,Gustavsson07b} These PAT effects are usually understood in terms of energy transfer between the QPC and the electrons/holes in the dot so that they could overcome the relevant energy barrier (Coulomb blockade or single-particle level spacing).~\cite{Aguado00} Therefore they are characterized by an energy cut-off corresponding to the mean level spacing of the dot, below which this energy transfer does not take place. Identifying the dense spectrum of our p-type QD as the source of the peculiar bias dependence of the tunnelling rates, it would be interesting to see if the detector has any back-action on the dot due to PAT and how much it contributes to the transport and its statistical properties.

Fig.\,\ref{fig:Fig4}(c) shows how the variation of the bias on the detector QPC ($V_{\rm QPC}^{\rm bias}$) influences the tunnelling rates of the dot. For small QPC bias $(V_{\rm QPC}^{\rm bias}<$ 70\,$\mu$V) detection is not possible due to low signal-to-noise ratio while for large QPC bias $(V_{\rm QPC}^{\rm bias}>$ 600\,$\mu$V) many fluctuators in the QPC are activated and the overall quality of the signal is degraded by the additional telegraph noise due to these fluctuators. The red shaded area shows the onset of this degradation. Fig.\,\ref{fig:Fig4}(d) shows the effect of QPC bias on the Fano factor and skewness of the hole transfer distribution. The agreement between the measurements and the two-state Markovian BN model (dashed line) implies that the effect of the QPC on the dot can be phenomenologically lumped into the tunnelling rates $\Gamma_{\rm in}$ and $\Gamma_{\rm out}$. This is in contrast to what is expected from a master equation calculation, which can be slightly modified to include the effect of PATs. It is shown in the Appendix using a simple model that the presence of an excited state generally alters the statistics obeyed by a dot unless relaxation is faster than both the tunnelling rates and the photo-excitation rate.

Furthermore, while $\Gamma_{\rm out}$ shown in Fig.\,\ref{fig:Fig4}(c) increases monotonically with the QPC bias, $\Gamma_{\rm in}$ decreases, suggesting that the influence of QPC on the dot is at least partially a simple gating effect. Considering the close proximity of the QPC leads and the dot tunnel barriers in Fig.\,\ref{fig:Fig1}(a) and the equal polarity of the dot and QPC biases, this is not surprising as most of the applied bias voltage drops over the QPC. As a result, the height of the source tunnelling barrier increases (decreasing $\Gamma_{\rm in}$) and the height of the drain tunnelling barrier decreases (increasing $\Gamma_{\rm out}$) for positive QPC bias. Moreover, since a symmetric bias of 700\,$\mu$V is applied to the dot (point A in Fig.\,\ref{fig:Fig2}(a)), $\Gamma_{\rm out}$ is expected to exhibit a step at a QPC bias of about 350\,$\mu$V as the holes can tunnel out to the source lead, while the measurement shows a monotonic increase of $\Gamma_{\rm out}$. Similar experiments were performed in an off-resonance configuration and no change in the tunnelling rates were observed for $V^{\rm bias}_{\rm QPC}<$~600\,$\mu$V. Therefore we conclude that the relaxation is dominant in our experiment and the results of Fig.\,\ref{fig:Fig4}(c)-(d) are mainly gating effects in the window of QPC biases investigated. For the other counting measurements the QPC bias was kept at 250\,$\mu$V.

\subsection{Normal vs. factorial cumulants}
\begin{figure}
\includegraphics[width=0.45\textwidth]{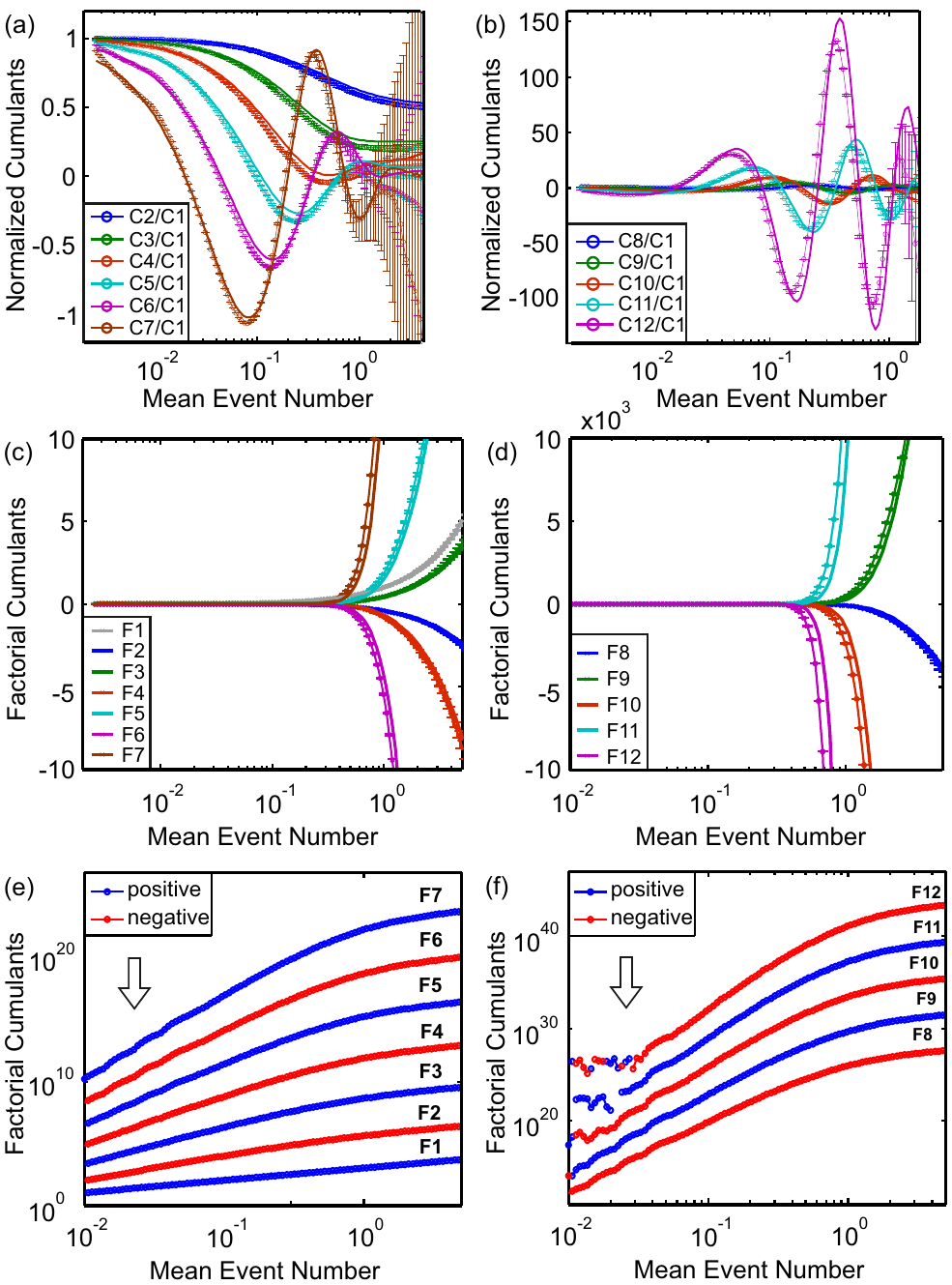} \caption{\label{fig:Fig5} \small{(a)-(b) First twelve normalized cumulants of charge transfer as a function of time (mean number of events in time traces with a given length cut from a very long time trace) calculated from the data collected at point A in Fig.\,\ref{fig:Fig2}(a) along with the predictions of BN model which fits remarkably well to the data. The amount of statistics decreases for longer time traces causing the size of error bar to increase. (c)-(d) First twelve factorial cumulants (FCs)~\cite{Kambly11} calculated from the normal cumulants. FCs of different order alternate sign and grow exponentially with time and therefore it is more convenient to follow their trend in logarithmic scale as shown in (e)-(f). No zero-crossing oscillations in FCs are observed consistent with a two-level Markovian model.~\cite{Kambly11} The white arrows point to the development of faint non-zero-crossing oscillations appearing in higher order FCs probably due to finite statistics.}}
\end{figure}

For a closer look at the statistics, we have calculated the first twelve cumulants of the tunnelling events and plotted them together with the predictions of the two-state Markovian BN model in Fig.\,\ref{fig:Fig5}(a)-(b). In general the finite bandwidth, the limited signal-to-noise ratio and the finite statistics influence the calculation of the cumulants. While the first two problems can be in principle taken into account by introducing additional Markovian states into the model,~\cite{Naaman06,Komijani11} the finite statistics is responsible for the error bars in Fig.\,\ref{fig:Fig5}(a)-(b). The latter is calculated from the covariance formula~\cite{Ferreira97} $\bracket{\Delta C_n \Delta C_m}=m!\sigma^{2m}\delta_{mn}N^{-1}+O(N^{-2})$. The $N$ in the denominator signifies the importance of the amount of statistics for a reasonable accuracy. For a fixed total number of events $K$ (two million in our case) used to calculate the cumulants as a function of $\bracket{n}$, the amount of statistics is equal to $N=K/\bracket{n}$. Also note that $C_2=\sigma^2$ eventually grows linearly with $\bracket{n}$ in the steady-state and therefore the error in the cumulant $C_m$ grows with $\bracket{n}^{(m+1)/2}$. Overall, a reasonable agreement between theory and experiment is obtained. Universal oscillations of the cumulants~\cite{Flindt09} highlight the difference between the distribution and a Gaussian distribution for which $C_n=0$ for $n>2$ and provide additional information about the probability distribution. An interesting piece of information is the position of the zeros of the generating function (ZGF) in the complex plane which, according to Abanov et al.~\cite{Abanov09}, is expected to be on the negative real axis for non-interacting systems. This is interesting as the strong interactions in p-type QDs may cause deviations from single-particle physics.~\cite{Komijani08} However, it is difficult to extract any useful information directly from normal cumulants as the poles of the cumulant generating function are displaced from the real axis by construction. Recently, Kambly et al.~\cite{Kambly11} proposed the use of factorial cumulants (FCs) for this purpose as any zero-crossing oscillations in the latter directly indicate the offset of ZGF from the real axis pointing towards relevance of interactions. We have calculated the first twelve FCs from our data which are shown in Fig.\,\ref{fig:Fig5}(c)-(d) and on the logarithmic scale in Fig.\,\ref{fig:Fig5}(e)-(f). Due to the logarithmic scaling of the FCs the latter plot is more convenient to follow the evolution of the results.~\cite{Kambly11} Note that consecutive FCs alternate sign as indicated by red and blue colors. For a two-state Markovian system no oscillations in the factorial cumulants are expected in agreement with the fact that there is no clear zero-crossing oscillations in the data. This again implies that the two-level system is a surprisingly good model to describe the statistical properties of our multi-level QD presumably due to the strong relaxation, as explained in the Appendix.

\section{Conclusion}

We have demonstrated time-averaged as well as time-resolved charge detection in a p-type GaAs QD. The extracted tunnelling rates suggest the presence of a dense spectrum of excited states in the dot contributing to transport. The full counting statistics of the QD is studied and shown to follow the two-state Markovian BN model in-spite of multi-level transport. This result and the absence of QPC back-action are interpreted in terms of a strong energy relaxation of holes in the QD which also ensures uni-directional transport within the bandwidth of our measurement.

\section*{Appendix}
Hidden states are a set of internal states of a system that are indistinguishable from the perspective of the charge detector. These could be some two-level systems in the barriers which affect the rates, some excited energy states within the dot or even some internal states of the charge detector.~\cite{Naaman06} In this Appendix we consider a simple model of the presence of an excited state in addition to the ground state of the dot, showing that when the hidden state is traced out, the dot appears to be obeying non-Markovian statistics. However, in the presence of strong relaxation, Markovian statistics is recovered. Similar problems have been considered by Belzig~\cite{Belzig05} and Flindt et al.~\cite{Flindt07} The starting point is the master equation for the three state system
\begin{equation}\label{eq:M3}
\left(\begin{array}{c}
\dot{p}_0\\
\dot{p}_g\\
\dot{p}_e
\end{array} \right)
=
\left(\begin{matrix}
-\Gamma^+_{\rm in} & z\Gamma^g_{\rm out} & z\Gamma^e_{\rm out} \\
\Gamma^g_{\rm in} & -\Gamma^g_{\rm out}-E^{eg} & \gamma^{ge} \\
\Gamma^e_{\rm in} & E^{eg} & -\gamma^{ge}-\Gamma^e_{\rm out} \\
\end{matrix}\right)
\left(\begin{array}{c}
p_0\\
p_g\\
p_e
\end{array} \right)
\end{equation}
where $\Gamma^{\pm}_{\rm in}=\Gamma^e_{\rm in}\pm\Gamma^g_{\rm in}$ and a similar expression for $\Gamma^{\pm}_{\rm out}$. $\gamma^{ge}$ is the relaxation rate, $E^{eg}$ is the rate of photon-assisted excitations and $z$ is the complex-value counting field~\cite{Kambly11} ($p_{i}=p_{i}(z,t)$ for $i=0,g,e$ and standard results are recovered for $z=1$). 
The charge detector is sensitive only to the occupation of the dot and not to the particular state occupied by the carrier. Therefore writing $p_1=p_g+p_e$, we have
\begin{equation}
\left(\begin{array}{c}
\dot{p}_0\\
\dot{p}_1\\
\dot{p}_{e}
\end{array} \right)
=
\left(\begin{matrix}
-\Gamma^+_{\rm in} & z\Gamma^g_{\rm out} & z\Gamma^-_{\rm out} \\
\Gamma^+_{\rm in} & -\Gamma^g_{\rm out} & -\Gamma^-_{\rm out} \\
\Gamma^e_{\rm in} & E^{eg} & -\gamma^{ge}-\Gamma^e_{\rm out} \\
\end{matrix}\right)
\left(\begin{array}{c}
p_0\\
p_1\\
p_e
\end{array} \right).
\end{equation}
Concentrating on the visible subspace by defining
\begin{equation}
{\bf w}\equiv 
\dot{{\bf v}}-
{\bf Mv}
\qquad
{\bf M}\equiv
\left(\begin{matrix} 
-\Gamma^+_{\rm in} & z\Gamma^g_{\rm out}\\
\Gamma^+_{\rm in} & -\Gamma^g_{\rm out}\\
\end{matrix}\right)
\end{equation}
where ${\bf v}\equiv \left(\begin{matrix} p_0 & p_1 \end{matrix}\right)^{T}$, it can be seen that deviations from Markovian statistics (${\bf w=0}$) are caused by the population of the excited state
\begin{equation}
{\bf w}=\Gamma^-_{\rm out}
\left(\begin{matrix}
z \\
-1
\end{matrix}\right)
p_e.
\end{equation}
This deviation is also proportional to $\Gamma^-_{\rm out}$ and it vanishes for $\Gamma^g_{\rm out}=\Gamma^e_{\rm out}$ as the two states become statistically indistinguishable. In the limit of $\gamma^{ge}\gg\ \Gamma^e_{\rm in},E^{eg}$, this population vanishes ($p_e\rightarrow 0$) in the steady-state, and the Markovian solution is recovered. This can be seen by using Eq.\,\ref{eq:M3} to eliminate $p_e$ from the previous equation
\begin{equation}
\dot{\bf w}+(\Gamma^e_{\rm out}+\gamma^{ge}){\bf w}=
\Gamma_{\rm out}^-\left(\begin{matrix}
z\Gamma^e_{\rm in} & E^{eg}\\
-\Gamma^e_{\rm in} & -E^{eg}
\end{matrix}\right)
{\bf v}
\end{equation}
In the limit $\gamma^{ge}\gg\ \Gamma^e_{\rm in},E^{eg}$, the right hand side can be neglected and ${\bf{w}}=0$ solves the resulting equation.\\

In the opposite limit of $\gamma,E\ra 0$, we expect that the tunnelling-out rate depends on the occupied state of the dot so that the histogram of $\Gamma_{\rm out}$ in Fig.\,\ref{fig:Fig1}(e) is no-longer a single exponential. Generally, the tunnelling-out (in) histogram will be a piece-wise linear function on a semi-log plot with the number of slopes equal to the number of excited states of $N+1$ ($N$)-charge configuration. Furthermore, the statistics will exhibit deviations from the two-level Markovian BN model shown here. We have never observed any deviation from the single-exponential distribution of the tunnelling rates and we attribute this to the dominant relaxation regime. The crossover regime in which the relaxation rate is finite but not enough to restore the Markovian statistics is beyond the scope of the present manuscript and we leave it as a future project.
\begin{acknowledgments}
Valuable discussions with B.~K\"ung, D.~Kambly, C.~Flindt are appreciated. We acknowledge M.~Csontos for his contributions during the beginning of this project and are grateful to the Swiss National Foundation via NCCR ``Quantum Science and Technology'' for financial support. D.~R.~and A.~D.~W.~acknowledge gratefully support from DFG SPP1285 and BMBF QuaHLRep 16BQ1035.
\end{acknowledgments}


%
%

\end{document}